\documentclass[aps,prd,preprintnumbers,showpacs,superscriptaddress,nofootinbib,amsmath,amssymb,floats,floatfix,showkeys,notitlepage,longbibliography]{revtex4-1}
\usepackage{comment}
\usepackage{graphicx}
\usepackage{subfigure}
\usepackage{palatino}
\usepackage[commandnameprefix=always]{changes}
\usepackage{hyperref}
\hypersetup{colorlinks=true,linkcolor=red,urlcolor=red,citecolor=red}
\usepackage[toc,page]{appendix}
\usepackage[normalem]{ulem}

\usepackage{lipsum}
\usepackage{graphicx}
\usepackage{subfigure}
\usepackage{palatino}
\usepackage{float}
\usepackage{sans}
\usepackage{adjustbox}
\usepackage{latexsym}
\usepackage{amsmath}
\usepackage{amssymb}
\usepackage{amsfonts}
\usepackage{dcolumn}
\usepackage{bm}
\usepackage{tikz}
\usepackage{bigints}
\usepackage{array,tabularx,multirow,booktabs}
\usepackage[tracking=true]{microtype}
\SetTracking{}{500}
\SetTracking{encoding={*}, shape=sc}{40}
\allowdisplaybreaks
\usepackage{adjustbox}
\usepackage{latexsym}
\usepackage{amsmath}
\usepackage{amssymb}
\usepackage{amsfonts}
\usepackage{dcolumn}
\usepackage{bm}
\usepackage{tikz}
\usepackage{bigints}
\usepackage{array,tabularx,multirow,booktabs}
\usepackage[tracking=true]{microtype}
\usepackage{color}
\allowdisplaybreaks

\begin{document}

\title{Mechanisms Behind the Aschenbach Effect in Non-Rotating Black Hole Spacetime}

\author{Mohammad Ali S. Afshar}
\email{m.a.s.afshar@gmail.com}
\affiliation{Department of Physics, Faculty of Basic Sciences,
University of Mazandaran\\
P. O. Box 47416-95447, Babolsar, Iran}
\affiliation{Canadian Quantum Research Center, 204-3002 32 Ave Vernon, BC V1T 2L7, Canada}
\author{Jafar Sadeghi}
\email{pouriya@ipm.ir}
\affiliation{Department of Physics, Faculty of Basic Sciences,
University of Mazandaran\\
P. O. Box 47416-95447, Babolsar, Iran}
\affiliation{Canadian Quantum Research Center, 204-3002 32 Ave Vernon, BC V1T 2L7, Canada}
\begin{abstract}
General relativity predicts that a rotating black hole drags the spacetime due to its spin. This effect can influence the motion of nearby objects, causing them to either fall into the black hole or orbit around it. In classical Newtonian mechanics, as the radius (r) of the orbit increases, the angular velocity ($\Omega$) of an object in a stable circular orbit decreases. However, Aschenbach discovered that for a hypothetical non-rotating observer, contrary to usual behavior, the angular velocity increases with radius in certain regions \cite{1}.  Although the possibility of observing rare and less probable ”rotational” behaviors in a  rotating structure is not unlikely or impossible. However, observing such behaviors in a "static" structure is not only intriguing but also thought-provoking, as it raises questions about the factors that might play a role in such phenomena. In seeking answers to this question, various static models, particularly in the context of nonlinear fields, were examined, with some results presented as examples in the article. Among the models studied, the model of Magnetic Black Holes in 4D Einstein–Gauss–Bonnet Massive Gravity Coupled to Nonlinear Electrodynamics (M-EGB-Massive) appears to be a candidate for this phenomenon. In the analysis section, we will discuss the commonalities of this model with previous models that have exhibited this phenomenon and examine the cause of this phenomenon. Finally, we will state whether this phenomenon is observable in other black holes and, if not, why.
\end{abstract}

\date{\today}

\maketitle
\section{Introduction}
The Aschenbach effect is a fascinating phenomenon that highlights the complex interplay between gravity, rotation, and the dynamics of matter near black holes. General relativity predicts that a rotating black hole drags spacetime around with it due to its spin. This effect can influence the motion of nearby objects, causing them to spiral in or orbit the black hole. Essentially, the curvature of spacetime induced by the black hole's rotation compels particles to begin orbiting the black hole, driven by the dynamics of the rotating spacetime.
In classical Newtonian mechanics, the angular velocity $\Omega$ of an object in a stable circular orbit decreases as the radius (r) of the orbit increases. This is because the gravitational force weakens with distance, requiring a lower orbital speed to maintain a stable orbit.
However, Aschenbach \cite{1} discovered that for a hypothetical observer who is locally non-rotating with respect to distant stars (a zero-angular-momentum observer, or ZAMO), the radial gradient of the angular velocity ($d\Omega/dr$) becomes positive in a radial range. This means that within this specific range, contrary to the usual behavior, the angular velocity increases with radius.
This phenomenon can have several intriguing implications:\\
1. Astrophysical Observations:\\
•  Quasi-Periodic Oscillations (QPOs): The Aschenbach effect can influence the frequencies of QPOs observed in the X-ray emissions from accreting black holes. These oscillations are thought to be linked to the motion of matter in the inner regions of the accretion disk, close to the black hole. The positive angular velocity gradient in the Aschenbach effect could lead to unique signatures in the QPOs, potentially providing a method to measure the spin of black holes with high precision \cite{1}.\\
2. Accretion Disk Dynamics:\\
•  Disk Instabilities: The change in the angular velocity gradient might affect the stability of the accretion disk. In regions where the angular velocity increases with radius, there could be new types of instabilities or resonances that could alter the flow of matter onto the black hole. This could impact the overall structure and behavior of the accretion disk \cite{2}.\\
3. Testing General Relativity:\\
•  Extreme Gravity Regimes: Observing the Aschenbach effect provides a way to test the predictions of general relativity in the strong-field regime. Since this effect is tied to the rapid spin of black holes, it offers a unique opportunity to study the behavior of spacetime under extreme conditions\cite{3}.\\
Although the study of such phenomena in rotating black holes is of great importance due to the structural rotation of the model and the frame-dragging effect of the surrounding spacetime, the occurrence of such phenomena has a high statistical probability for these black holes. However, it becomes even more intriguing when we ask whether such phenomena can also occur in non-rotating black holes.
In response to this question, "Wei" and "Liu" demonstrated in their study that simpler models like Schwarzschild and Reissner-Nordström do not exhibit such characteristics. However, in the relatively more complex Dyonic model, the possibility and likelihood of this phenomenon forming exist \cite{4}. Subsequently, "Pavan Kumar Yerra" and colleagues examined this in a massive charged model, and their results also indicated the presence of this phenomenon \cite{5}.
However, what common factor existed in the two previous black holes that could lead to this seemingly unlikely phenomenon in non-rotating black holes? To answer this question, we examined various black hole models, including those with linear and nonlinear fields, string theory, dark matter, and dark energy, both in simple and combined forms. We observed that although this phenomenon could exist in most of these models, it was not physically observable for various reasons. During these studies, we arrived at a model that incorporates the Gauss-Bonnet term, massive gravity, and a magnetic field in its action \cite{6}. This model appears to be a candidate for demonstrating the Aschenbach effect, which we will discuss in the analysis section, focusing on the characteristics of this model and its commonalities with previous models. This shared factor may explain the occurrence of this phenomenon in non-rotating black holes.
But, before that, to fully understand this phenomenon, we must first take a comprehensive look at the classification of space based on the location of the photon sphere, which plays a crucial role in this context, and conduct a precise analysis of the spatial positions of TCOs (Timelike Circular Orbits) \cite{7}.\\
Accordingly, in this paper, Section 2 will provide a very brief introduction to the method we will use to study photon  spheres and TCOs, as thoroughly explained in several previous papers. In Section 3, we will introduce the model and  study and classify the space around the model in terms of photon spheres and TCOs. Section 4 we will examine the  effects and compare the commonalities of this model with previously studied models, exploring the reasons for the  occurrence of this phenomenon, and we state whether this  phenomenon exists in other black holes and, if so, why it is not observable. In Section 5, we will summarize our findings in the conclusion.
\section{Methodology}
To study the photon sphere, we utilize the Charges calculation and topological method. Since this method is well-explained in \cite{8} and our previous works \cite{9,10,11}, we will refrain from delving deeply into its concepts and instead refer interested readers to these works.
Additionally, in our examination of TCOs , we use the relationships established in the studies conducted in \cite{7,12,13}. Due to the comprehensive nature of these articles, we will only highlight the fundamental relationships and main concepts.
\subsection{Topological Photon Sphere } 
\begin{figure}[H]
 \begin{center}
 \subfigure[]{
 \includegraphics[height=6.5cm,width=6cm]{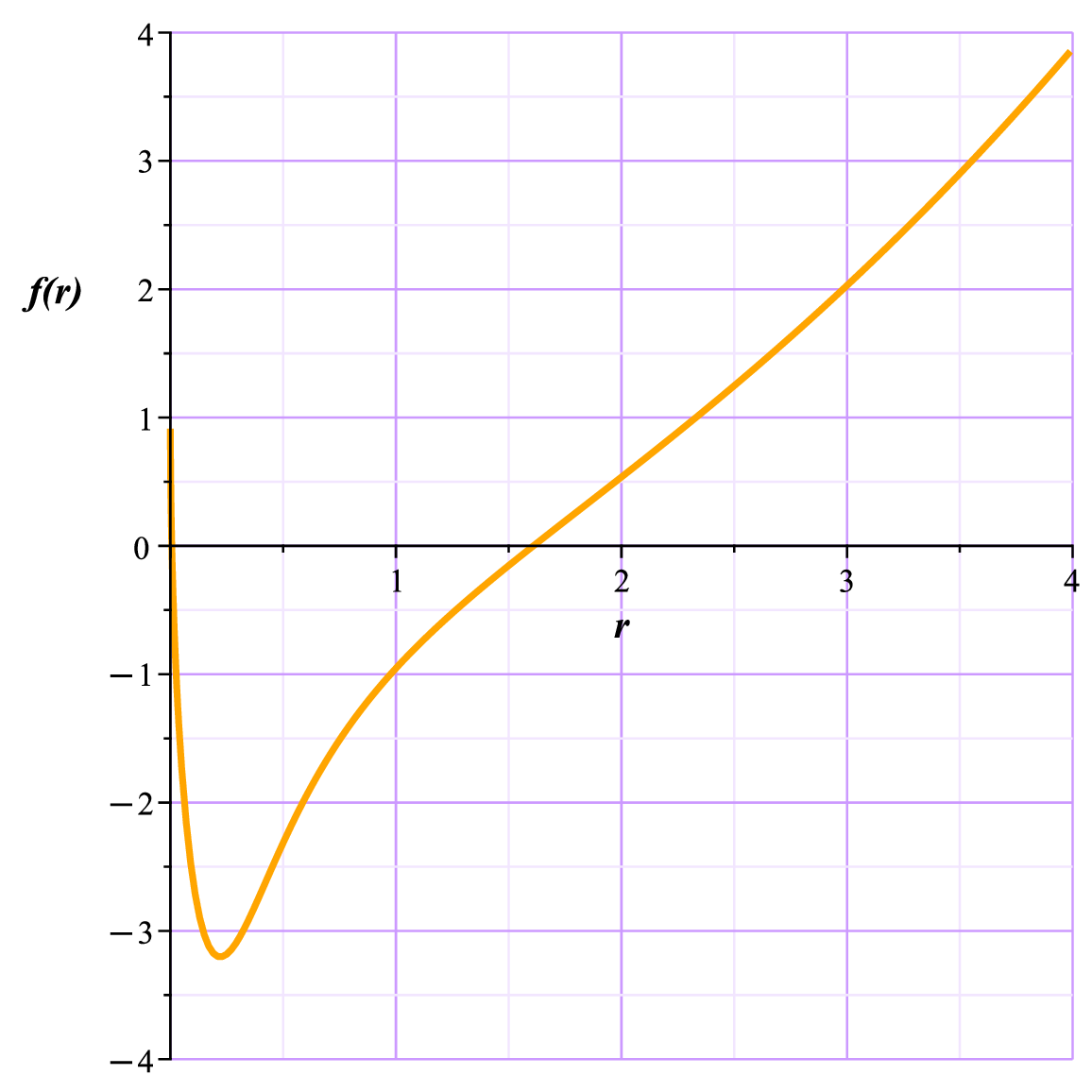}
 \label{1a}}
 \subfigure[]{
 \includegraphics[height=6.5cm,width=6cm]{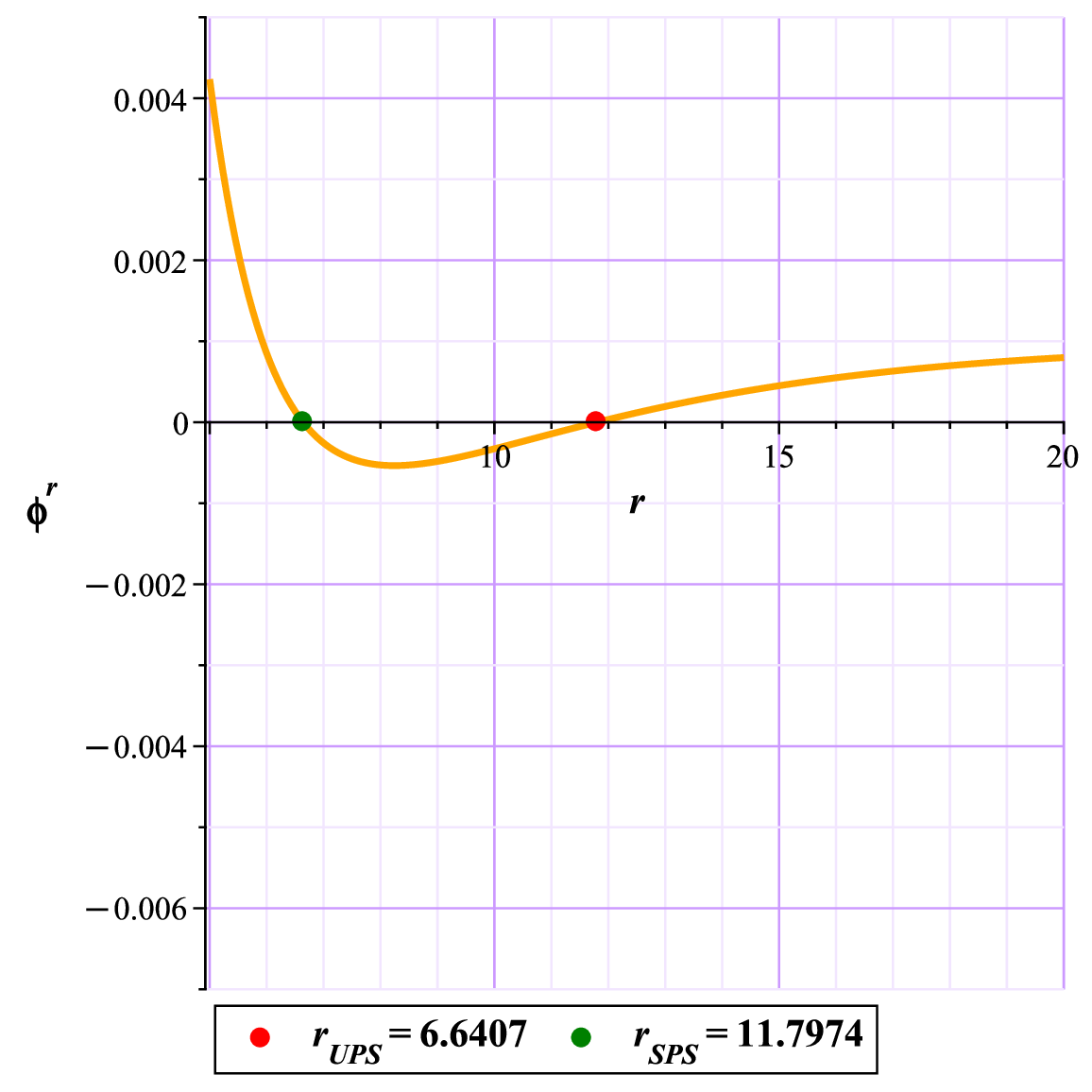}
 \label{1b}}
\subfigure[]{
 \includegraphics[height=6.5cm,width=6cm]{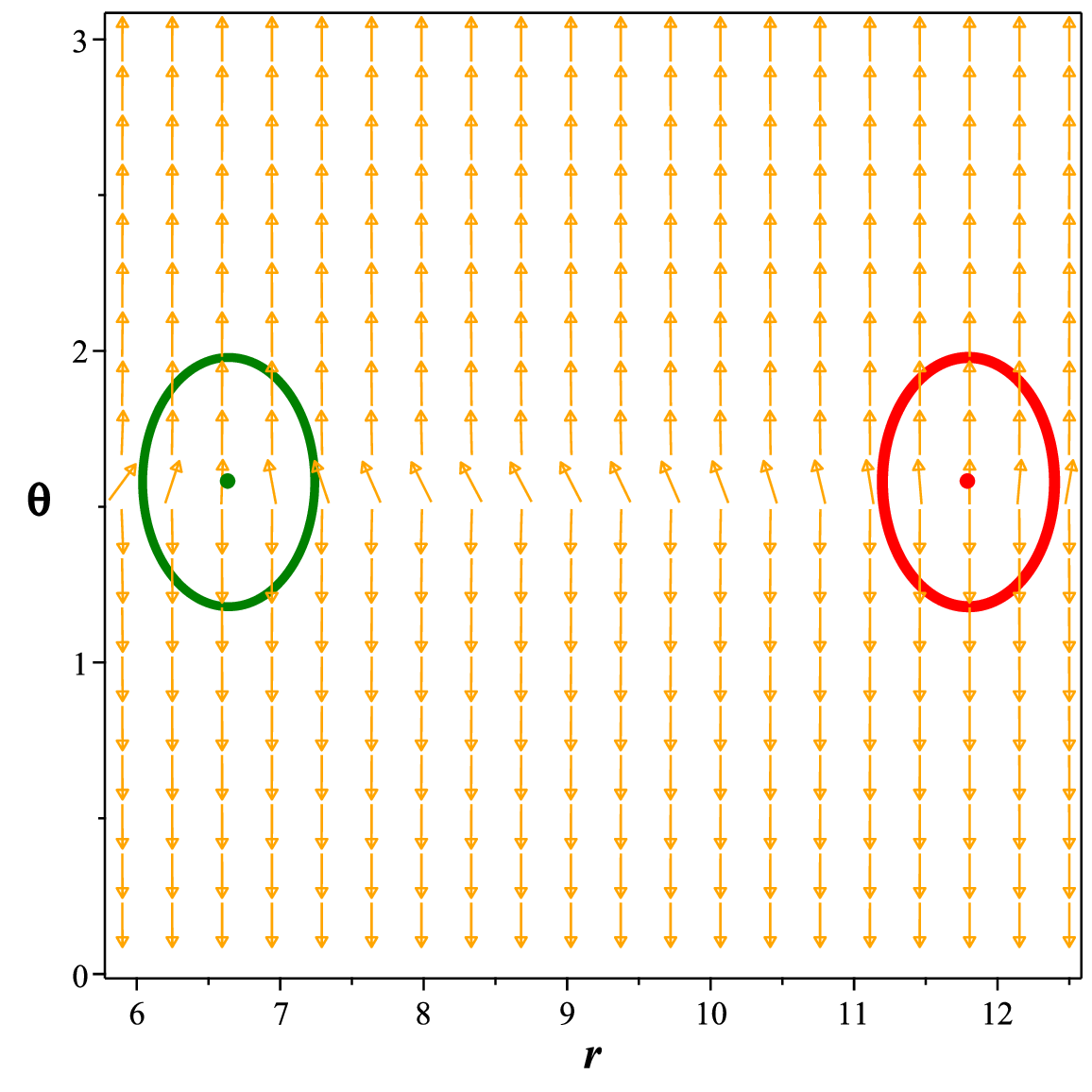}
 \label{1c}}
 \subfigure[]{
 \includegraphics[height=6.5cm,width=6cm]{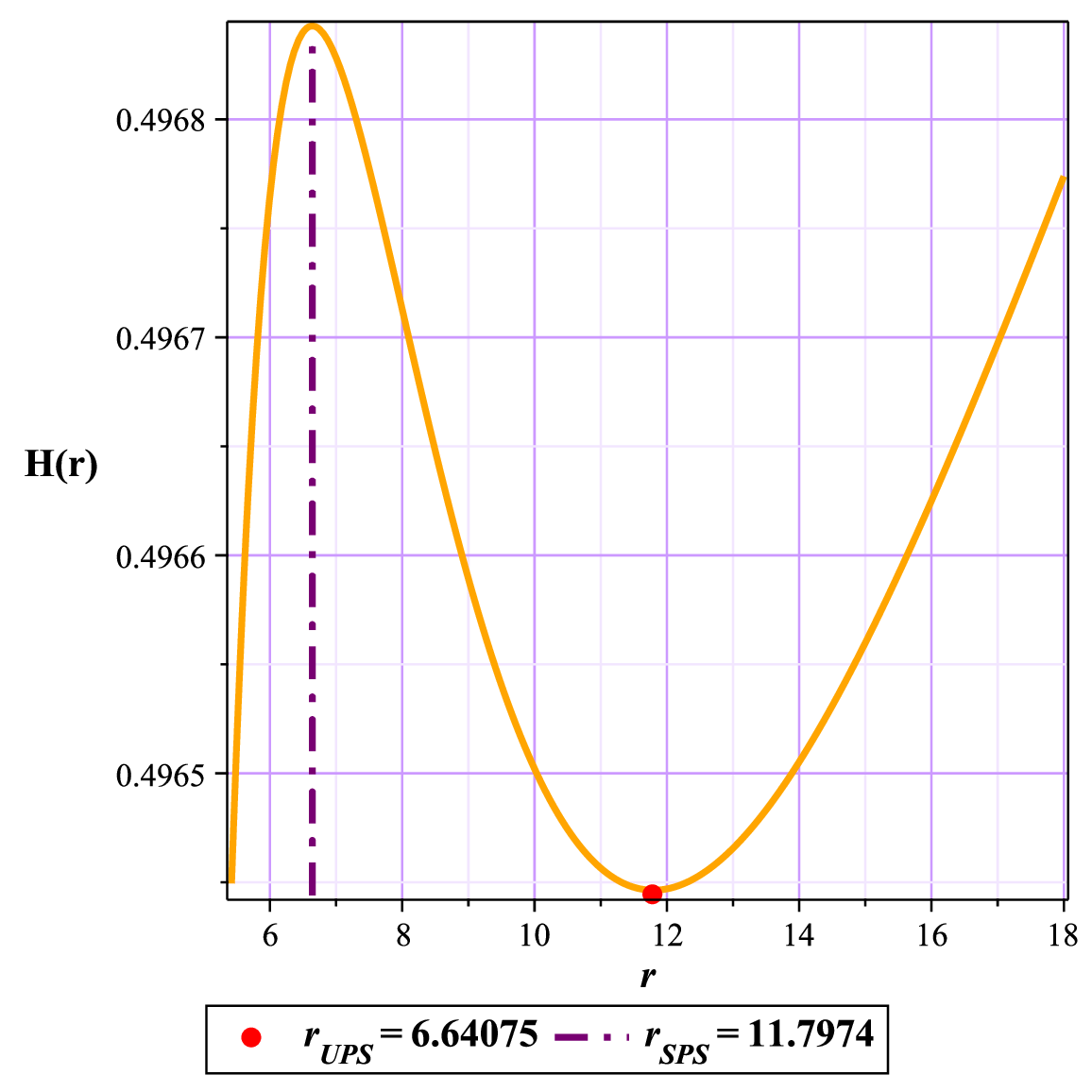}
 \label{1d}}
   \caption{\small{Fig (1a): Metric function with $l = 2, \beta = 1.5, c = 1, C_{1} = -1, C_{2} = 1, M = 0.8, q = 2, \alpha = 0.01, m = 0.5$ for  MEGB Massive BH , (5b): $\phi^{r}$ function, The Unstable photon spheres(UPS) are located at $ r=(6.64075)$ The Stable photon spheres(SPS) are located at $ r=(11.79744)$, (1c): The UPS with topological charge -1 located at (r, $\theta$) = (6.64075, 1.57) and SPS with topological charge +1 located at (r, $\theta$) = (11.79744, 1.57) in the (r ,$\theta$) plane of the normal vector field n (1d): the topological potential }}
 \label{1}
\end{center}
\end{figure}

First, we consider a general vector field as  $\phi$  which can be decomposed into two components, $\phi^r$ and $\phi^\theta$,\cite{8,9,10,11}
\begin{equation}\label{1}
\phi=(\phi^{r}, \phi^{\theta}),
\end{equation}
 We can  rewrite this vector as $\phi=||\phi||e^{i\Theta}$, where $||\phi||=\sqrt{\phi\cdot\phi}$, or $\phi = \phi^r + i\phi^\theta$.
Based on this, the normalized vector can be defined as,
 \begin{equation}\label{2}
n^\sigma=\frac{\phi^\sigma}{||\phi||},
\end{equation}
 where $\sigma=1,2$  and  $(\phi^1=\phi^r)$ , $(\phi^2=\phi^\theta)$.
With respect to the necessity of spherical symmetry as a prerequisite for studying this method, and given the most general form of the metric in 4-dimensional form we have:
\begin{equation}\label{(3)}
\mathit{ds}^{2}=-\mathit{dt}^{2} f \! \left(r \right)+\frac{\mathit{dr}^{2}}{g \! \left(r \right)}+\left(d\theta^{2}+d\varphi^{2}
\sin \! \left(\theta \right)^{2}\right) h \! \left(r \right)=\mathit{dr}^{2} g_{\mathit{rr}}-\mathit{dt}^{2} g_{\mathit{tt}}+d\theta^{2} g_{\theta \theta}+d\varphi^{2} g_{\varphi \varphi},
\end{equation}
Now the new form of effective potential, can be written in the form below \cite{8,9,10,11}:
\begin{equation}\label{(4)}
\begin{split}
H(r, \theta)=\sqrt{\frac{-g_{tt}}{g_{\varphi\varphi}}}=\frac{1}{\sin\theta}\bigg(\frac{f(r)}{h(r)}\bigg)^{1/2}.
\end{split}
\end{equation}
In terms of H(r), the vector field vector field $\phi=(\phi^r,\phi^\theta)$ can be written as follows:
\begin{equation}\label{(5)}
\begin{split}
&\phi^r=\frac{\partial_rH}{\sqrt{g_{rr}}}=\sqrt{g(r)}\partial_{r}H,\\
&\phi^\theta=\frac{\partial_\theta H}{\sqrt{g_{\theta\theta}}}=\frac{\partial_\theta H}{\sqrt{h(r)}}.
\end{split}
\end{equation}
Before concluding this section, we recommend that researchers refer to references \cite{8,9,10} for a deeper understanding of the calculation and study of topological charge. However, in a brief overview, it must be stated that, from a classical perspective, the potential must always be a function of the existing configuration in the study space and should not depend on the physical conditions of the incoming particles. Accordingly, the effective potential derived from the Lagrangian of black hole action can be rewritten to meet this prerequisite, which leads to Eq. (\ref{(4)}).
Now, based on this new effective potential, if we form the vector field $\phi$ Eq. (\ref{(5)}), in other words, by mapping the potential to a two-dimensional space $(r, \theta)$ using Duan’s $\phi-mapping$ method, the optima of the H function, which represent the locations of photon spheres \cite{10}, act as zero points in the new space and cause changes in the behavior of field lines around them. To provide a simple mental image, in this scenario, the field lines around the local maxima of the potential (unstable photon spheres) behave like negative electric charges, whereas around the local minima (stable photon spheres), they behave like positive charges, as clearly shown in Fig. (\ref{1c}).\\
Now, using the above statements and considering the definition of topological charges, each photon sphere can be assigned a charge. It is important to note that in our papers \cite{9,10}, we demonstrated that when the model structure is in the form of a black hole, typically with the necessary condition of having  event horizon, it often exhibits an unstable photon sphere outside the horizon with a total topological charge of -1. In terms of the potential diagram, this implies the appearance of at least one maximum in the spacetime outside the horizon.
Conversely, when the structure lacks a horizon (i.e., a naked singularity), it usually presents, in addition to the unstable photon sphere with a topological charge of -1, one or two stable photons spheres with a topological charge of +1 in the studied spacetime, which, from the potential perspective, correspond to stable minima. These are very important points that we will refer to again in the future.
\subsection{Time-like Circular Orbits (TCOs) }
The geodesics of any gravitational model are deeply dependent on the shape of the potential resulting from the fields involved in the model's action structure. These geodesics can either lack observable turning points or possess them, forming loops. In null geodesics, these loops are known as photon spheres, while in timelike geodesics, they are referred to as TCOs. Both types of orbits can be stable or unstable.
Our assumptions in this study include a static and axisymmetric spacetime with  $\mathbb{Z}_2$ symmetry in a 1+3 dimensional framework. Given these symmetries, discussing the equatorial plane does not compromise the generality of the analysis.
Based on the metric equation ,Eq. (\ref{(3)}), We consider the following quantities \cite{7}:
\begin{equation}\label{(6)}
A =g_{\varphi \varphi} E^{2}+g_{\mathit{tt}} L^{2},
\end{equation}
\begin{equation}\label{(7)}
B =-g_{\varphi \varphi} g_{\mathit{tt}},
\end{equation}
which E,L are the energy and the angular momentum. Now the Lagrangian can be recast as:
\begin{equation}\label{(8)}
2 \mathfrak{L} =-\frac{A}{B}=\zeta,
\end{equation}
where $\zeta = -1, 0 $ for time-like, null  geodesics, respectively. Using the above Lagrangian, the effective potential can be rewritten as follows:
\begin{equation}\label{(9)}
V_{\mathit{eff} \! \left(\zeta \right)}\! \left(r \right)=\zeta +\frac{A}{B}.
\end{equation}
The concept of angular velocity (as measured by an observer at infinity) and $\xi$ in terms of metric parameters.
\begin{equation}\label{(10)}
\Omega_{\pm}=\frac{g_{\mathit{tt}} L}{g_{\varphi \varphi} E},
\end{equation}
\begin{equation}\label{(11)}
\xi_{\pm}=-A \! \left(r^{\mathit{cir}},\Omega_{\pm},\Omega_{\pm}\right),
\end{equation}
in which $\pm$ is a sign of prograde/retrograde orbits \cite{7}.Although the main focus here is not on TCOs, it is worthwhile to provide a brief explanation. However, interested readers who seek a deeper understanding of these relations and more examples are encouraged to refer to \cite{7,12,13}.
Given the definition of effective potential Eq. (\ref{(9)}) , we know that for having a circular orbit, a returning potential is needed, which is obtained by setting the potential equation to zero. As stated, this results in the formation of circular orbits, which are referred to as photon spheres for photons and massless particles, and as TCO (Timelike Circular Orbits) for massive particles, which can be either stable or unstable.
From classical mechanics, we know that to study stability, it is sufficient to examine the second derivative of the effective potential, and the sign of this second derivative determines the stability or instability of the orbit \cite{7,12,13}. Considering the continuity of the potential function, we expect a boundary between these two categories of orbits, i.e., stable and unstable TCOs, to exist. This boundary is known as the Marginally Stable Circular Orbit (MSCO), and it represents the smallest stable circular orbit.\\
An important aspect to consider in the study of TCOs (Timelike Circular Orbits) is the value of the $\xi$ function. In regions where $\xi$ is negative, the energy and angular momentum become imaginary, rendering these areas effectively forbidden for the presence of TCOs, which we will talk more about this in the future. 
\section{Magnetic black holes in 4D Einstein–Gauss–Bonnet massive gravity coupled to nonlinear electrodynamics}
Massive gravity black hole models offer a rich and complex framework for exploring gravitational phenomena. They provide new solutions and insights that challenge and extend our understanding of black holes, cosmology, and the fundamental nature of gravity. One of the primary motivations for developing massive gravity theories was to explain the accelerated expansion of the universe without invoking dark energy. By giving the graviton (the hypothetical quantum particle that mediates the force of gravity) a small mass, researchers hoped to modify gravity on cosmological scales, potentially explaining this acceleration \cite{14}. These theories can provide alternative descriptions of black holes that might avoid some of the singularities predicted by general relativity. This is particularly important for understanding the true nature of black holes and the fundamental structure of spacetime \cite{15}.
They allow for the existence of new black hole solutions, including those with "hair" (additional parameters beyond mass, charge, and angular momentum). These solutions can provide deeper insights into the nature of black holes and their interactions with surrounding matter and fields\cite{15}. Massive gravity theories introduce new degrees of freedom, which can affect the stability of black holes\cite{15}. Studying these stability properties helps in understanding the robustness of black hole solutions and their potential observational signatures.\\
The AdS D-dimensional action for M-EGB-Massive black hole is given by \cite{6}:
\begin{equation*}\label{0}
S=\int d^{D}x\frac{(R -2 \Lambda +\frac{ \alpha}{\mathrm{D}-4}\mathcal{G}+\mathcal{L}_{\mathit{NED}}+ m^{2}\sum \mathcal{U}_{i} c_{i}) \sqrt{-g}}{16 \pi},
\end{equation*}
where R is Ricci scalar, $\Lambda$ is the cosmological constant, $\alpha$ is Gauss–Bonnet coupling parameter, $\mathcal{G}$ is the Gauss–Bonnet term, m is a parameter related to graviton mass, $c_{i}$ is
constant, $\mathcal{U}_{i}$ are symmetric polynomials of eigenvalues of matrix $\mathcal{K^\mu}_{\nu}=\sqrt{g^{\mu\alpha}h_{\alpha\nu}}$ which more precise calculations can be found in different references \cite{6,6.1} and $\mathcal{L}_{\mathit{NED}}$ is defined as follows:
\begin{equation*}\label{(0)}
\mathcal{L}_{\mathit{NED}} =-\frac{2 q^{2}}{2 r^{\mathrm{D}-2} \sqrt{\beta}\, q +r^{2 \mathrm{D}-4}},
\end{equation*}
where $\beta$ is the positive coupling. The 4D-metric for such a black hole that we consider in the exact form is \cite{6}:
\begin{equation*}\label{(0)}
f_{0}=\frac{\arctan \! \left(\frac{r \sqrt{2}}{2 \sqrt{\sqrt{\beta}\, q}}\right) q^{2} \sqrt{2}}{2 r^{3} \sqrt{\sqrt{\beta}\, q}}-\frac{1}{l^{2}}-\frac{\left(2 C_{2} c^{2}+r C_{1} c \right) m^{2}}{2 r^{2}},
\end{equation*}
\begin{equation}\label{(12)}
f =1+\frac{\left(1-\sqrt{1+4 \left(\frac{2 M}{r^{3}}+f_{0}\right) \alpha}\right) r^{2}}{2 \alpha}
\end{equation}
where $ l^2=-3/\Lambda$ is AdS radius, $\alpha$ the is Gauss–Bonnet coupling parameter, M is related to the mass of the black hole, q is the magnetic charge.
\subsection{Topological Photon Sphere }
Considering the metric function Eq. (\ref{(3)}) and also according to the following equations:
\begin{equation}\label{(13)}
f \! \left(r \right)=g \! \left(r \right),
\end{equation}
\begin{equation}\label{(14)}
h \! \left(r \right)=r,
\end{equation}
and with respect to Eq. (\ref{(4)}), Eq. (\ref{(5)}) we have:
\begin{equation}\label{(15)}
H =\frac{\sqrt{4+\frac{2 \left[1-\sqrt{1+4 \left(\frac{2 M}{r^{3}}+f_{0}\right) \alpha}\right] r^{2}}{\alpha}}}{2 \sin \! \left(\theta \right) r}.
\end{equation}
\begin{equation*}\label{(0)}
\varphi_{1}=\frac{\frac{\arctan \left(\frac{r \sqrt{2}}{2 \sqrt{\sqrt{\beta}\, q}}\right) q^{2} \sqrt{2}\, \alpha  l^{2}}{4}+\sqrt{\sqrt{\beta}\, q}\, \left\{\left[\left(-\frac{1}{2} c^{2} m^{2} C_{2} r -\frac{1}{4} c \,m^{2} C_{1} r^{2}+M \right) \alpha +\frac{r^{3}}{8}\right] l^{2}-\frac{\alpha  r^{3}}{2}\right\}}{\sqrt{\sqrt{\beta}\, q}\, r^{3} l^{2}},
\end{equation*}
\begin{equation*}\label{(0)}
\varphi_{2}=q \left(-\frac{1}{3} c^{2} m^{2} C_{2} r -\frac{1}{12} c \,m^{2} C_{1} r^{2}+M \right) \sqrt{\beta}+\frac{\left(-\frac{1}{3} c^{2} m^{2} r^{2} C_{2}-\frac{1}{12} c \,m^{2} r^{3} C_{1}+M r -\frac{1}{6} q^{2}\right) r}{2},
\end{equation*}
\begin{equation*}\label{(0)}
\varphi_{3}=-\frac{2 r \sqrt{\sqrt{\beta}\, q}\, \left(\sqrt{\beta}\, q +\frac{r^{2}}{2}\right) \sqrt{2}\, \sqrt{\varphi_{1}}}{3}+\frac{\sqrt{2}\, q^{2} \left(\sqrt{\beta}\, q +\frac{r^{2}}{2}\right) \arctan \! \left(\frac{r \sqrt{2}}{2 \sqrt{\sqrt{\beta}\, q}}\right)}{4}+\varphi_{2} \sqrt{\sqrt{\beta}\, q},
\end{equation*}
\begin{equation}\label{(16)}
\phi^{r}=\frac{3 \varphi_{3} \csc \! \left(\theta \right) \sqrt{2}}{4 \sqrt{\varphi_{1}}\, \sqrt{\sqrt{\beta}\, q}\, r^{3} \left(\sqrt{\beta}\, q +\frac{r^{2}}{2}\right)}.
\end{equation}
\begin{equation}\label{(17)}
\phi^{\theta}=-\frac{\sqrt{4+\frac{2 \left[1-\sqrt{1+4 \left(\frac{2 M}{r^{3}}+f_{0}\right) \alpha}\right] r^{2}}{\alpha}}\, \cos \! \left(\theta \right)}{2 \sin \! \left(\theta \right)^{2} r^{2}}
\end{equation}
Given the extensive parametric diversity of the aforementioned model, an analytical solution is exceedingly challenging. Consequently, we resorted to numerical methods. Additionally, as we were pursuing a specific objective, which will be elaborated upon in subsequent sections, a wide array of combinations could have been selected to achieve our goal. Here, we will examine two such combinations.

\begin{center}
\textbf{$\alpha = 0.01$ }
\end{center}
\begin{center}
\textbf{$\alpha = 0.85$ }
\end{center}
\begin{figure}[H]
 \begin{center}
 \subfigure[]{
 \includegraphics[height=6.5cm,width=8cm]{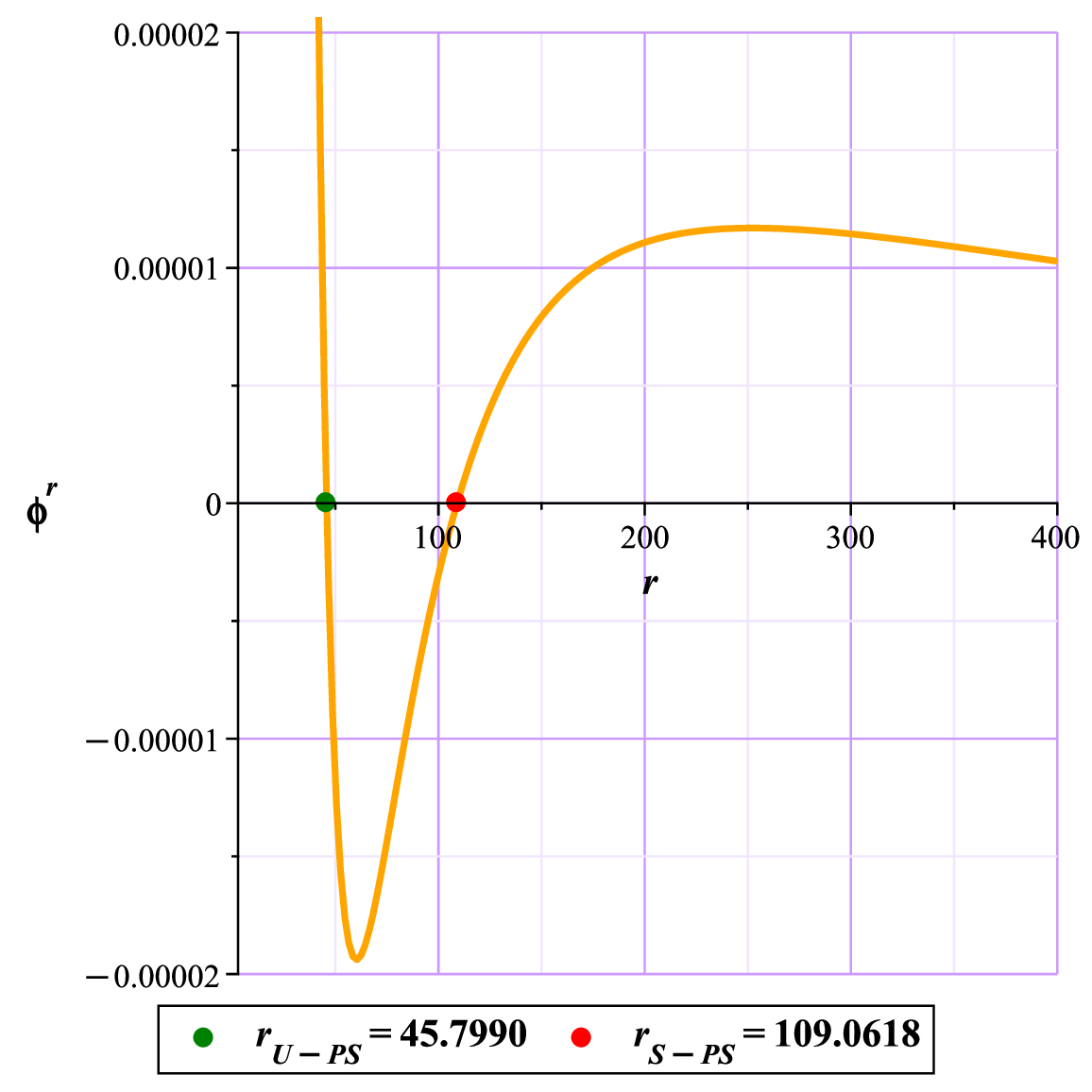}
 \label{2a}}
 \subfigure[]{
 \includegraphics[height=6.5cm,width=8cm]{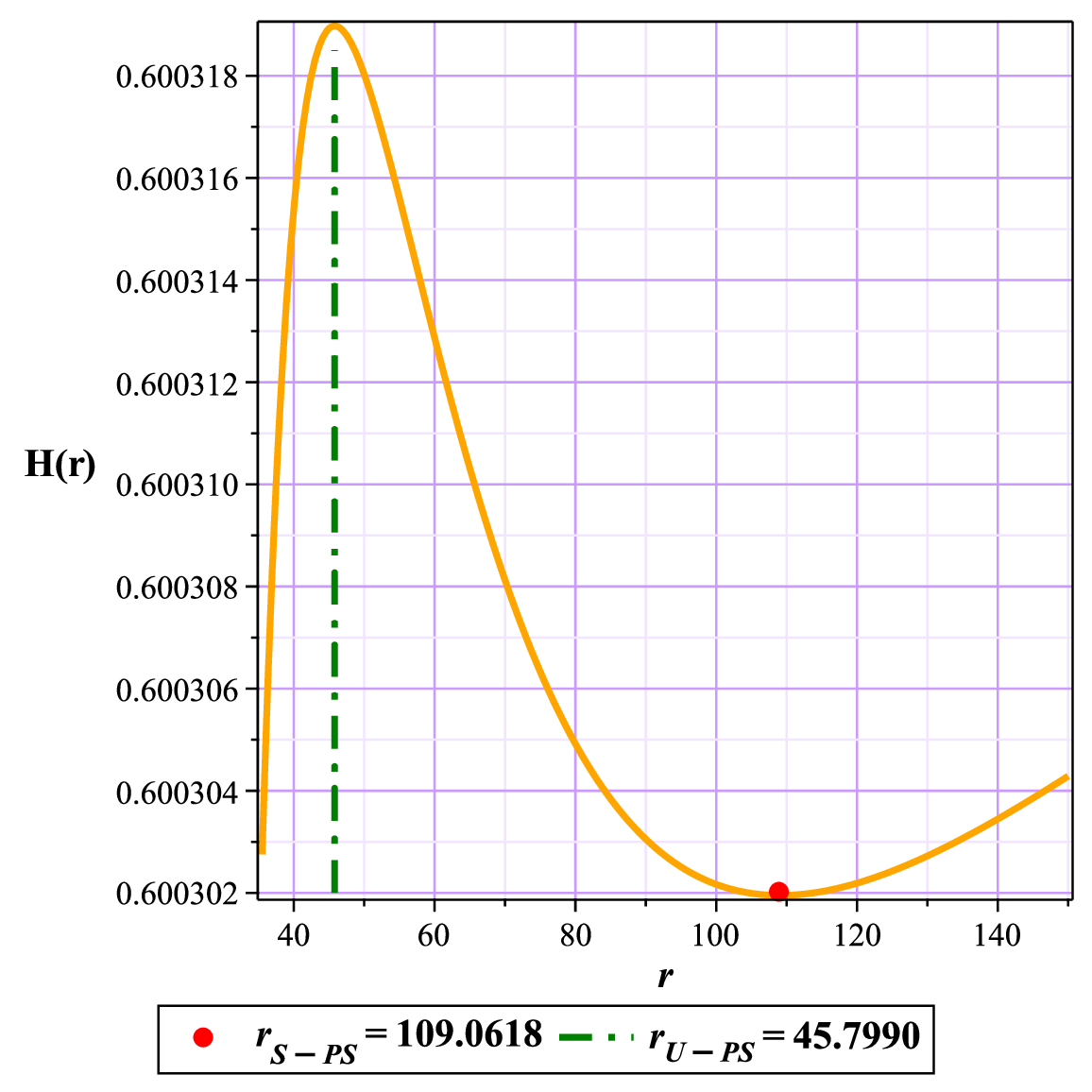}
 \label{2b}}
   \caption{\small{Fig (2a): $\phi^{r}$ function, U-PS are located at $ r=(45.79903)$ The S-PS are located at $ r=(109.06185)$ for $l = 2, \beta = 8, c = 1, C_{1} = -1, C_{2} = 1, M = 1, q = 5, \alpha = 0.85, m = 0.1$, (2b): the topological potential H(r) for  M-EGB-Massive BH  }}
 \label{2}
\end{center}
\end{figure}
The photon spheres of the black hole under study are illustrated in Fig. (\ref{1}) and Fig. (\ref{2}) for two different parametric conditions. What distinguishes the photon sphere in this model from other models studied using topological methods \cite{9,10,11,12,13,16,17,19} is that, in cases where the model represents a black hole with an event horizon, it typically shows an unstable photon sphere outside its horizon (a local potential maximum). Conversely, when the model lacks an event horizon and represents a naked singularity, stable photon spheres (local minima in the effective potential diagram) generally appear in the spacetime under study. Here, we observe that despite maintaining the event horizon, the structure exhibits both a stable photon sphere and an anti-photon sphere outside the horizon. This is a point that warrants closer examination of its implications.\\Before concluding this section, it is worthwhile to address an important question:\\ Among the terms added to the action that led to the construction of this model, 
which ones, considering their respective parameters, have played a more significant role in the emergence of this stable photon sphere outside the event horizon, distinguishing this model from others?\\
To answer this question, it is better to examine the radial component of the $\phi$ function, which represents the location of the photon spheres, in the limiting cases created by the elimination of parameters.
Accordingly, in the first step, for the limit $\beta \rightarrow 0$ we could see that the term  $\mathcal{L}_{\mathit{NED}}$ change to general form of Maxwell electrodynamics and the model transform to the magnetically charged AdS black hole in 4D EGB massive gravity with the metric as follows \cite{6,20}:
\begin{equation*}\label{(0)}
f =1+\frac{\left(1-\sqrt{1+4 \left[\frac{2 M}{r^{3}}+\frac{q^{2}}{r^{4}}-\frac{1}{l^{2}}-\frac{\left(2 C_{2} c^{2}+r C_{1} c \right) m^{2}}{2 r^{2}}\right] \alpha}\right) r^{2}}{2 \alpha}.
\end{equation*} 
If we apply the limit $\alpha \rightarrow 0$, the model becames a 4D massive Einstein gravity black holes with NED, whose metric will be as follows \cite{6,21}:
\begin{equation*}\label{(0)}
f =1-\frac{2 M}{r}-\frac{\arctan \! \left(\frac{r \sqrt{2}}{2 \sqrt{\sqrt{\beta}\, q}}\right) q^{2} \sqrt{2}}{2 \sqrt{\sqrt{\beta}\, q}\, r}+\frac{r^{2}}{l^{2}}+\left(c^{2} C_{2}+\frac{1}{2} c r C_{1}\right) m^{2}.
\end{equation*}
and finally if we choose  the limit $m \rightarrow 0$  for the massless model we have a 4D EGB massless gravity black holes with NED, which the metric for this model could be written in the form \cite{6}:
\begin{equation*}\label{(0)}
f =1+\frac{\left(1-\sqrt{1+4 \left[\frac{2 M}{r^{3}}+\frac{\arctan \left(\frac{r \sqrt{2}}{2 \sqrt{\sqrt{\beta}\, q}}\right) q^{2} \sqrt{2}}{2 r^{3} \sqrt{\sqrt{\beta}\, q}}-\frac{1}{l^{2}}\right] \alpha}\right) r^{2}}{2 \alpha}.
\end{equation*}
\begin{figure}[H]
 \begin{center}
 \subfigure[]{
 \includegraphics[height=7cm,width=8cm]{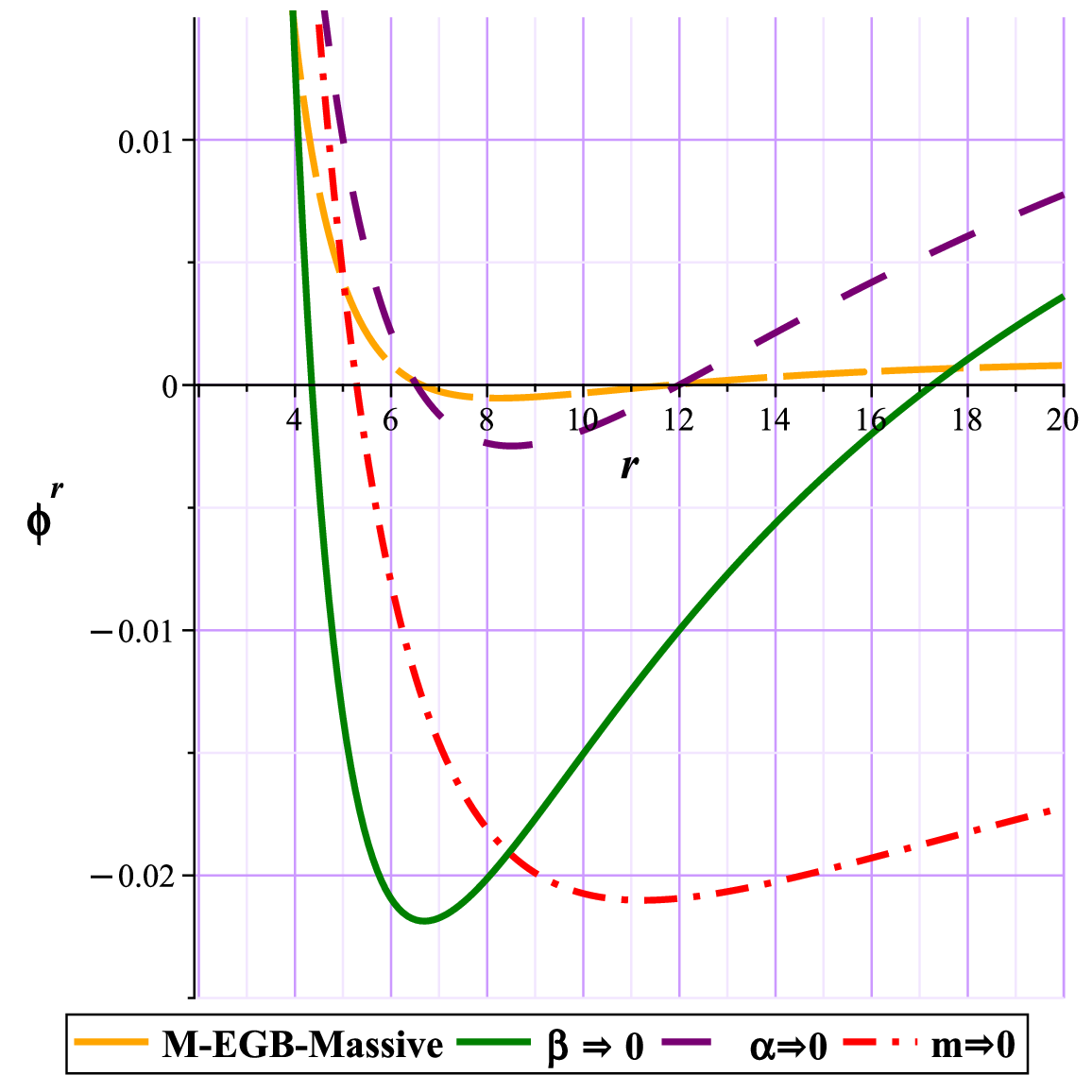}
 \label{3.1a}}
 \caption{\small{Fig (3): $\phi^{r}$ VS r for different limit states of  M-EGB-Massive in black hole with respect to $l = 2, \beta = 1.5, c = 1, C_{1} = -1, C_{2} = 1, M = 0.8, q = 2, \alpha = 0.01, m = 0.5$}}
 \label{3.1}
\end{center}
\end{figure}
As illustrated in  Fig. (\ref{3.1}), we have plotted the radial component of the function $\phi$ in a primary state (orange dashed line) and three limit states as described. It is evident that the emergence of a stable photon sphere outside the event horizon appears to result from the combination of the electromagnetic (whether linear or nonlinear) with the mass component of the graviton in the action of the model, referred to as Massive.
Since, Our investigations have shown that the G-B model, in different form, actually does not exhibit such conditions. Additionally, in Fig. (\ref{3.1}), the red dash-dot line, which corresponds to the presence of only the NED component, also indicates the presence of only an unstable photon sphere. Therefore, it can be concluded that as long as we remain in the massive form, the conditions for the emergence of a stable photon sphere beyond the event horizon seem to be met.

\subsection{TCOs}
With respect to Eq. (\ref{(6)}) and Eq. (\ref{(7)}) and Eq. (\ref{(11)}) for this model we will have:
\begin{equation}\label{(18)}
A =E^{2} r^{2}-\left[1+\frac{\left(1-\sqrt{1+4 \left(\frac{2 M}{r^{3}}+f_{0}\right) \alpha}\right) r^{2}}{2 \alpha}\right] L^{2}
\end{equation}
\begin{equation}\label{(19)}
B =\left[1+\frac{\left(1-\sqrt{1+4 \left\{\frac{2 M}{r^{3}}+f_{0}\right\} \alpha}\right) r^{2}}{2 \alpha}\right] r^{2}
\end{equation}
\begin{equation*}\label{(0)}
\xi_{1}=-\frac{2 r \sqrt{\sqrt{\beta}\, q}\, \left(\sqrt{\beta}\, q +\frac{r^{2}}{2}\right) \sqrt{2}\, \sqrt{\varphi_{1}}}{3}+\frac{\sqrt{2}\, q^{2} \left(\sqrt{\beta}\, q +\frac{r^{2}}{2}\right) \arctan \! \left(\frac{r \sqrt{2}}{2 \sqrt{\sqrt{\beta}\, q}}\right)}{4}
\end{equation*}
\begin{equation*}\label{(0)}
\xi_{2}=q \left(-\frac{1}{3} c^{2} m^{2} C_{2} r -\frac{1}{12} c \,m^{2} C_{1} r^{2}+M \right) \sqrt{\beta}+\frac{\left(-\frac{1}{3} c^{2} m^{2} r^{2} C_{2}-\frac{1}{12} c \,m^{2} r^{3} C_{1}+M r -\frac{1}{6} q^{2}\right) r}{2}
\end{equation*}
\begin{equation}\label{(20)}
\xi =-\frac{3 \sqrt{2}\, \left(\xi_{1}+\sqrt{\sqrt{\beta}\, q}\, \xi_{2}\right)}{4 \sqrt{\sqrt{\beta}\, q}\, \sqrt{\varphi_{1}}\, r \left(\sqrt{\beta}\, q +\frac{r^{2}}{2}\right)}
\end{equation}
\begin{figure}[H]
 \begin{center}
 \subfigure[]{
 \includegraphics[height=7cm,width=8cm]{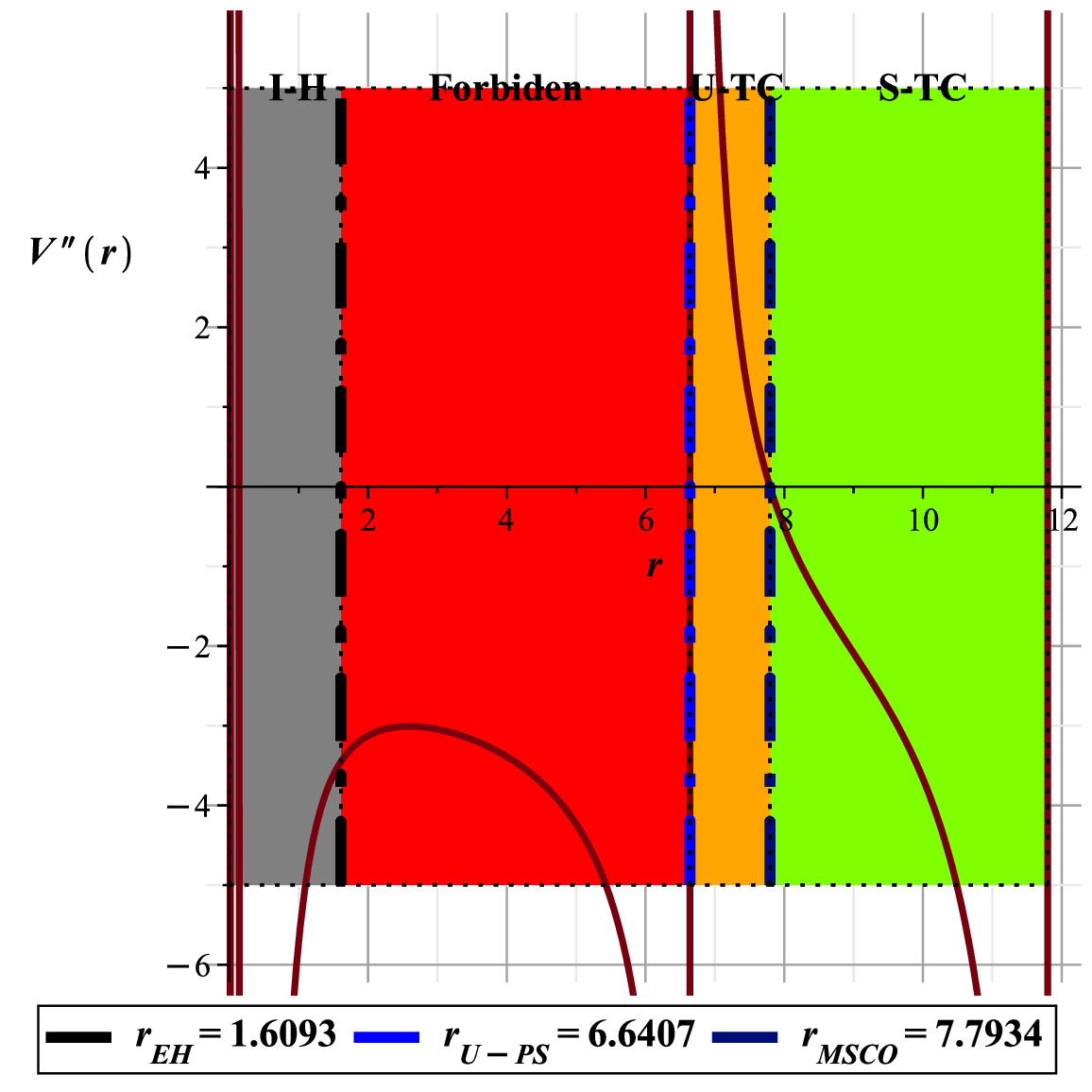}
 \label{3a}}
 \caption{\small{ MSCO localization and space classification for the M-EGB-Massive in black hole mode}}
 \label{3}
\end{center}
\end{figure}
It is noteworthy that the distribution of TCOs in the black hole models we have studied, as mentioned in \cite{7}, typically follows a pattern where UTCOs appear after U-PS  and connect to STCOs after crossing the MSCO (Marginally Stable Circular Orbit) boundary \cite{7,12,13}. However, in this case Fig. (\ref{3}), due to the emergence of a stable photon sphere, the spacetime behavior appears to differ, resembling that of a naked singularity. Consequently, TCOs are distributed around the pair of photon spheres \cite{7,12,13}.
\section{Discussion and analysis of Aschenbach effect}
\begin{figure}[H]
\begin{center}
\subfigure[]{
\includegraphics[height=7cm,width=7cm]{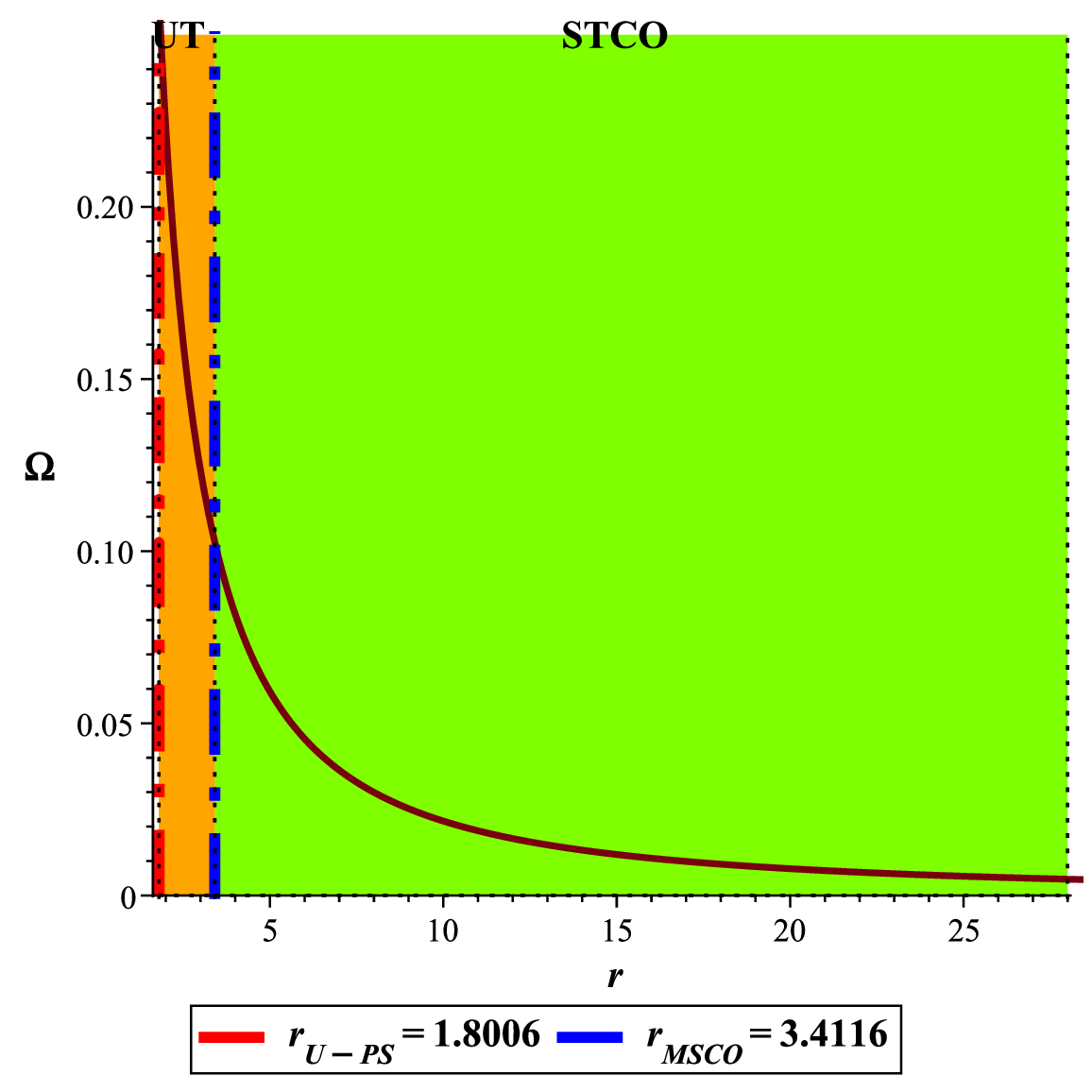}
\label{4a}}
\subfigure[]{
\includegraphics[height=7cm,width=7cm]{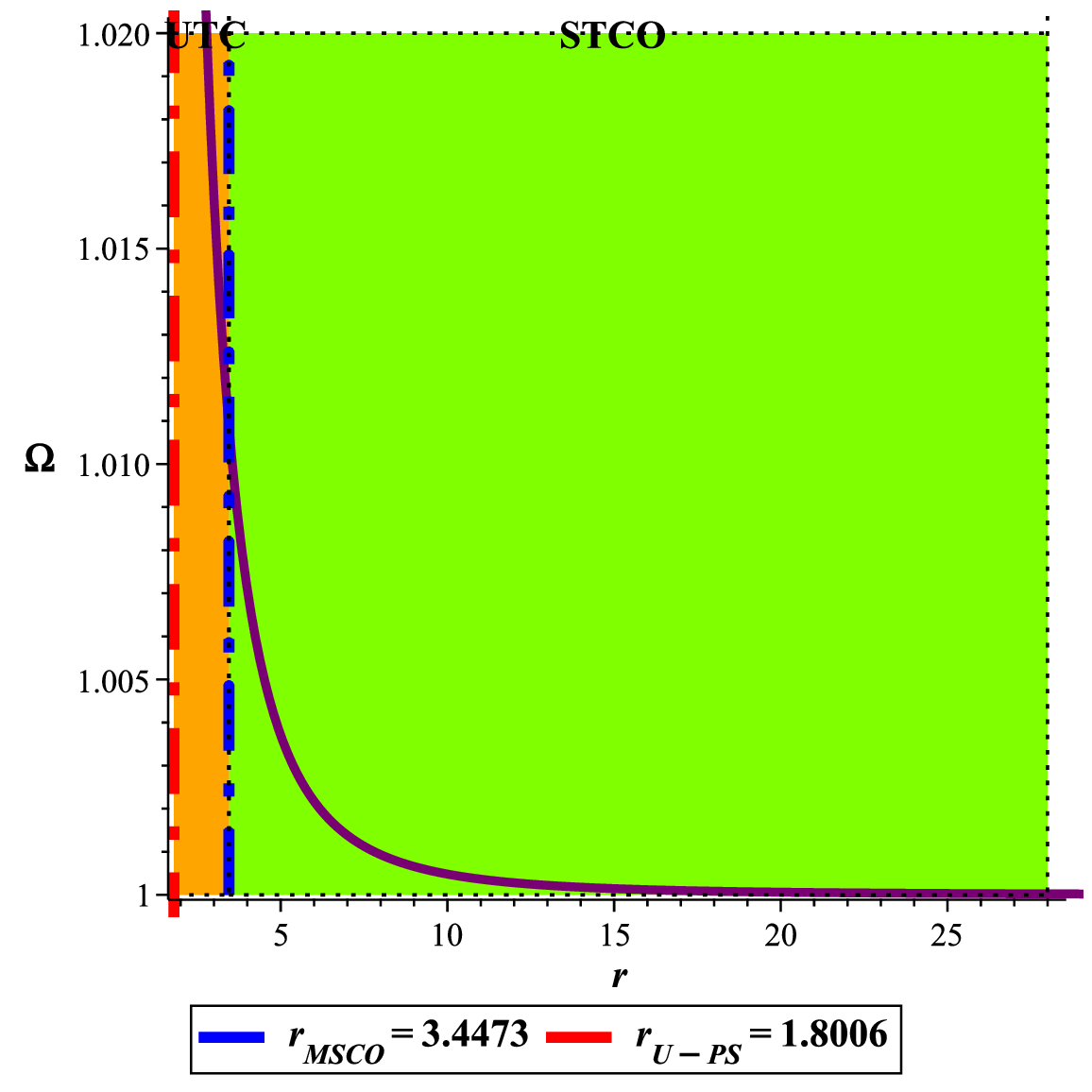}
\label{4b}}
\subfigure[]{
\includegraphics[height=7cm,width=7cm]{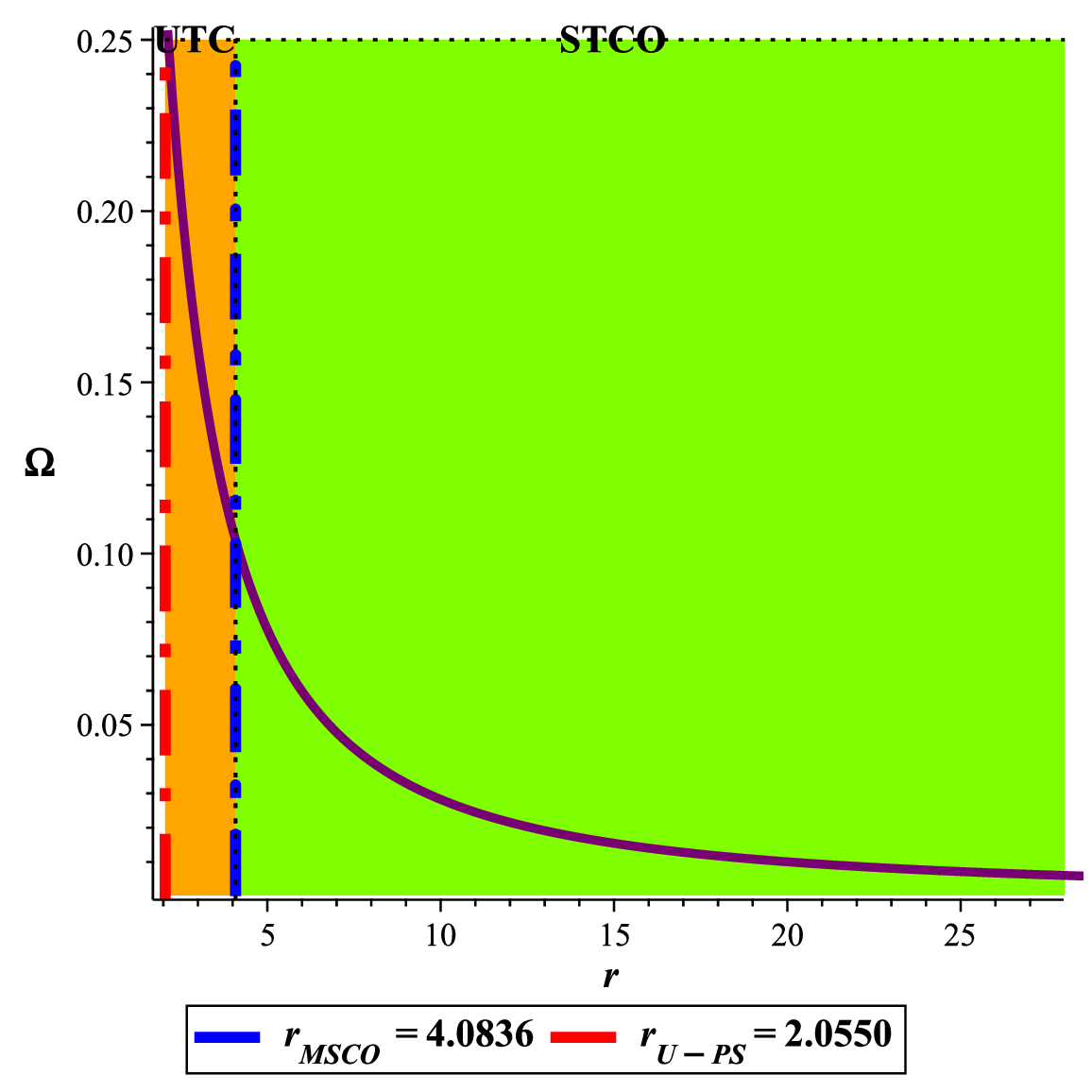}
 \label{4c}}
   \caption{\small{ Angular velocity VS r for Fig (5a): Charged Four-Dimensional Gauss-Bonnet Black Hole with Strings Cloud and Non-Commutative Geometry, Fig (5b): Non-Commutative Einstein-Born-Infeld black holes, Fig (5c): Euler-Heisenberg black hole surrounded by perfect fluid dark matter, \cite{12,13} }}
 \label{4}
\end{center}
\end{figure}

\begin{figure}[H]
 \begin{center}
 \subfigure[]{
 \includegraphics[height=5.5cm,width=8cm]{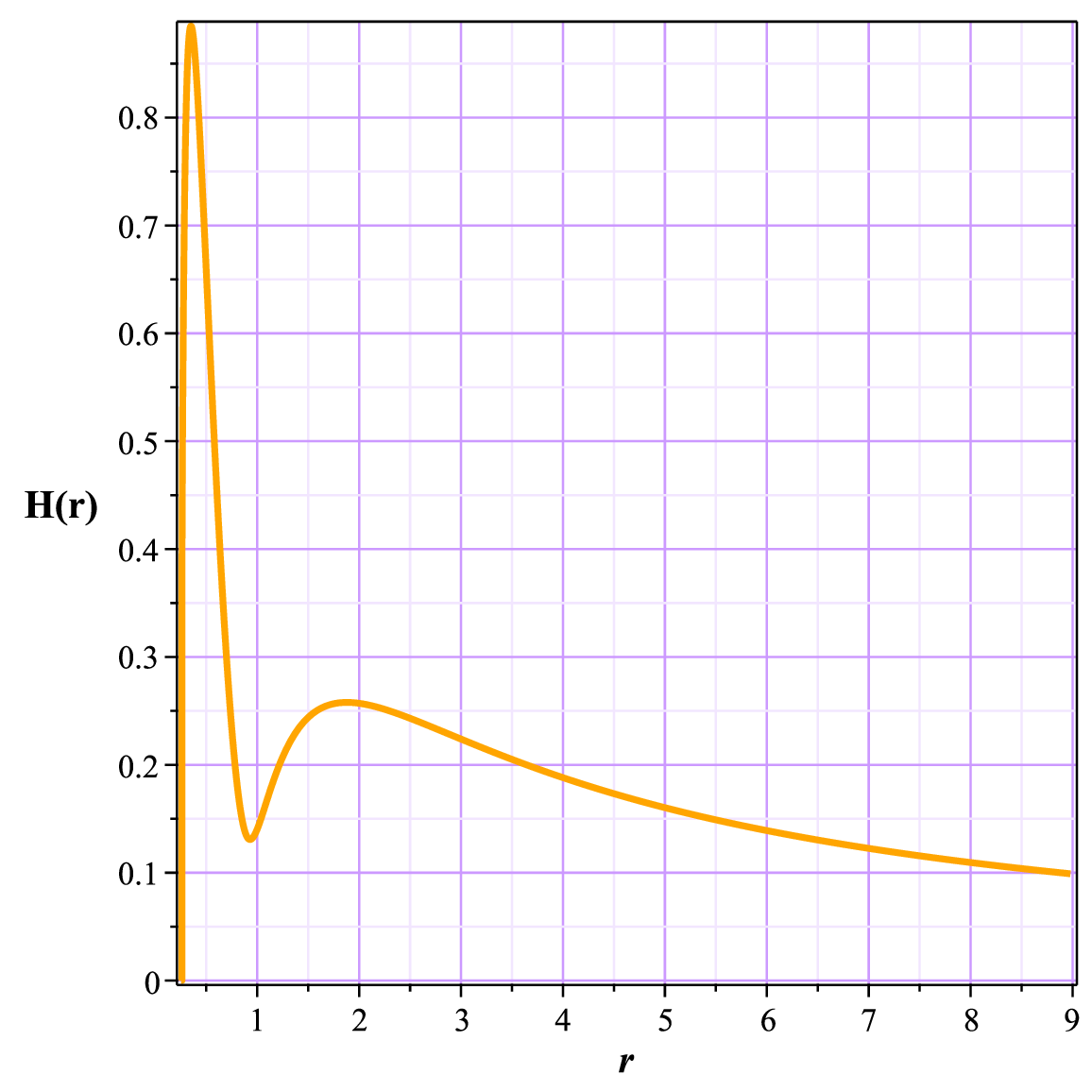}
 \label{5a}}
 \caption{\small{ the topological potential H(r) for Dyonic BH, \cite{4}}}
 \label{5}
\end{center}
\end{figure}

\begin{figure}[H]
 \begin{center}
 \subfigure[]{
 \includegraphics[height=6.5cm,width=8cm]{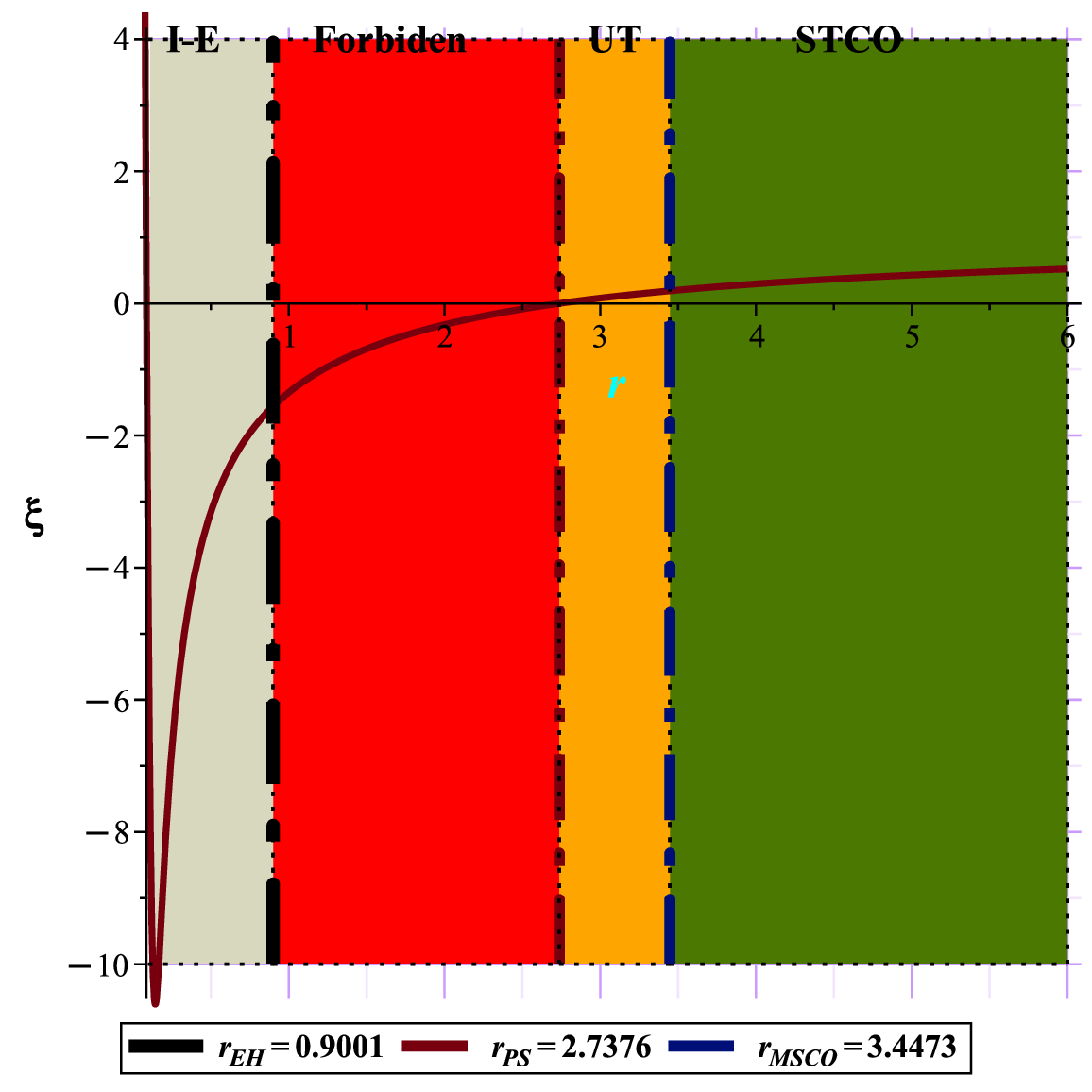}
 \label{7a}}
 \subfigure[]{
 \includegraphics[height=6.5cm,width=8cm]{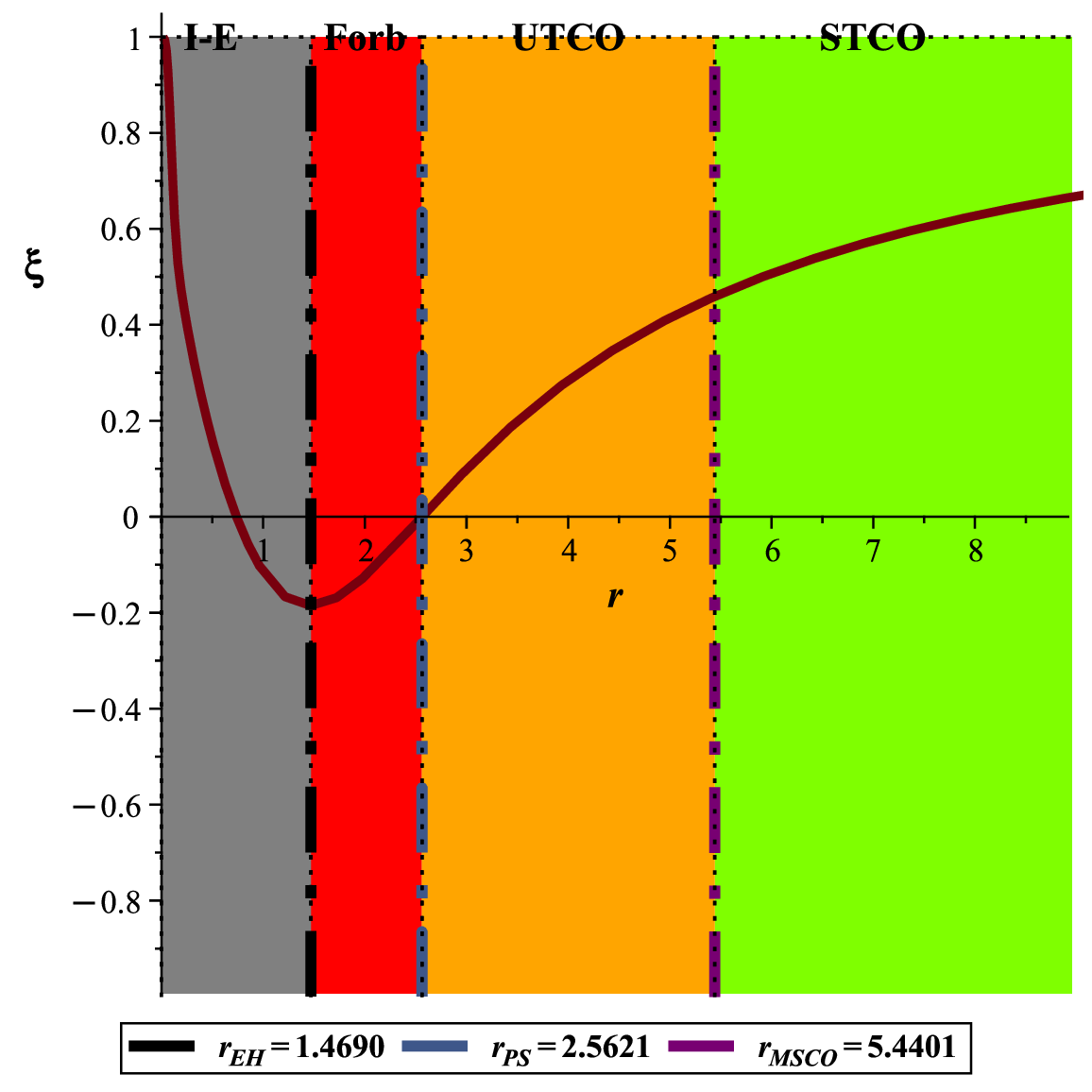}
 \label{7b}}
   \caption{\small{Fig (8a): $\xi$ diagram for Non-Commutative Einstein-Born-Infeld black hole , (8b): $\xi$ diagram for Non-Commutative 4D E-G-B black hole, \cite{12,13}}}
 \label{7}
\end{center}
\end{figure}

\begin{figure}[H]
 \begin{center}
 \subfigure[]{
 \includegraphics[height=6.5cm,width=8cm]{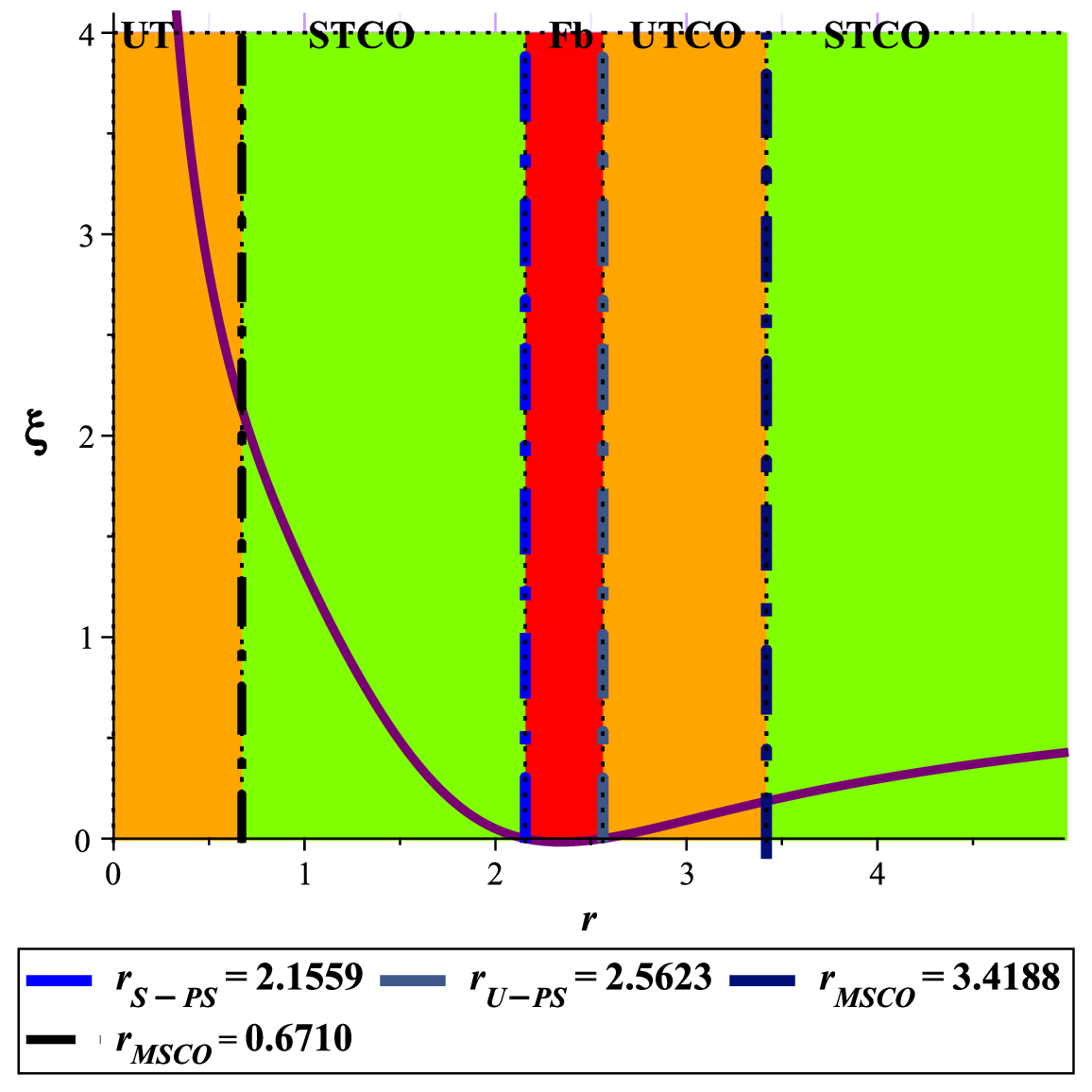}
 \label{8a}}
 \subfigure[]{
 \includegraphics[height=6.5cm,width=8cm]{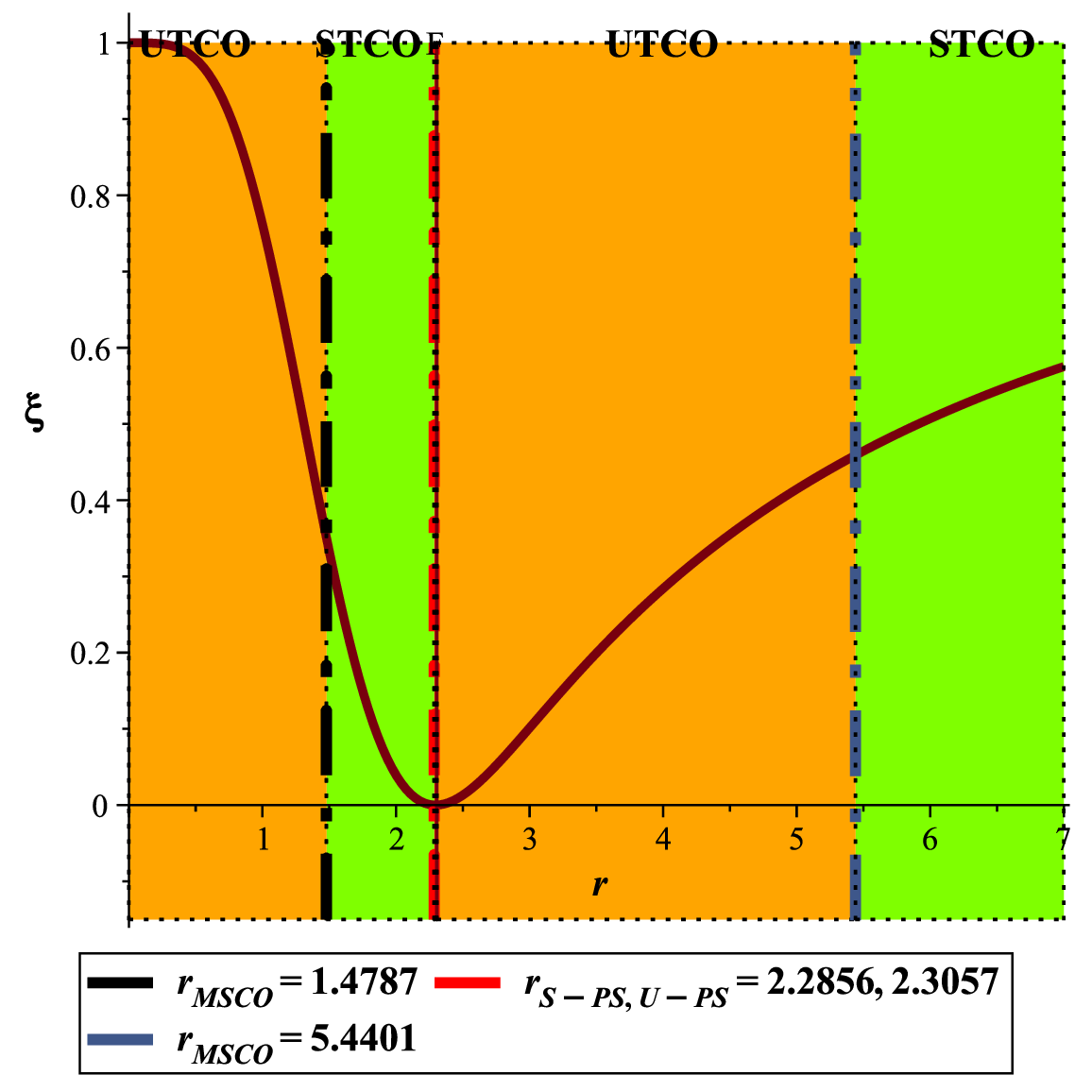}
 \label{8b}}
   \caption{\small{Fig (9a): $\xi$ diagram for Non-Commutative Einstein-Born-Infeld naked singularity , (9b): $\xi$ diagram for Non-Commutative 4D E-G-B naked singularity, \cite{12,13}}}
 \label{8}
\end{center}
\end{figure}

As mentioned in the introduction, the possibility of observing rare and less probable "rotational" behaviors in a rotating structure is not unlikely or impossible \cite{1,18}. However, observing such behaviors in a static structure is not only intriguing but also thought-provoking, as it raises questions about the factors that might contribute to such phenomena. In pursuit of answers to this question, we tried to study various static models, especially in the form of nonlinear fields. But before anything else, for this study, we must first introduce the equation of angular velocity \cite{7,12,13}.
\begin{equation}\label{(21)}
\begin{split}
&\Omega=\frac{L f}{E \,r^{2}},\\
&E_{\pm}=\frac{f}{\sqrt{\xi_{\pm}}},\\
&L_{\pm}=-\frac{r^{2} \Omega_{\pm}}{\sqrt{\xi_{\pm}}},\\
\end{split}
\end{equation}
in which ± is a sign of prograde/retrograde orbits \cite{7}. As previously mentioned, most models we have studied in our prior work using topological photon spheres typically exhibit a U-PS outside the event horizon in the form of a black hole. In this scenario, UTCOs originate from the photon sphere boundary and connect to STCOs after passing through the MSCO. "Wei" and "Liu" demonstrated in their work \cite{4} that Schwarzschild and Reissner-Nordström black holes exhibit such behavior. Also, Our studies show that the Aschenbach effect does not occur in many different forms of static black holes, as we have shown a few examples in Fig. (\ref{4}). They also demonstrated that the specific Dyonic model exhibited the Aschenbach effect \cite{4}. Upon examining the photon sphere of that Dyonic model, we observe a common feature with our work: the emergence of a stable photon sphere outside the horizon,Fig. (\ref{5}). 
As can be seen in the Fig. (\ref{5}), two U-PS and one S-PS have appeared outside the horizon in the potential diagram.
Now, if we examine the photon spheres calculated by us Fig. (\ref{1}),Fig. (\ref{2}), we clearly see that our black hole also exhibits this characteristic. This is a point we emphasized in the photon sphere section. The question now is whether our black hole model will also exhibit the Aschenbach effect.
\begin{figure}[H]
 \begin{center}
 \subfigure[]{
 \includegraphics[height=6.5cm,width=9cm]{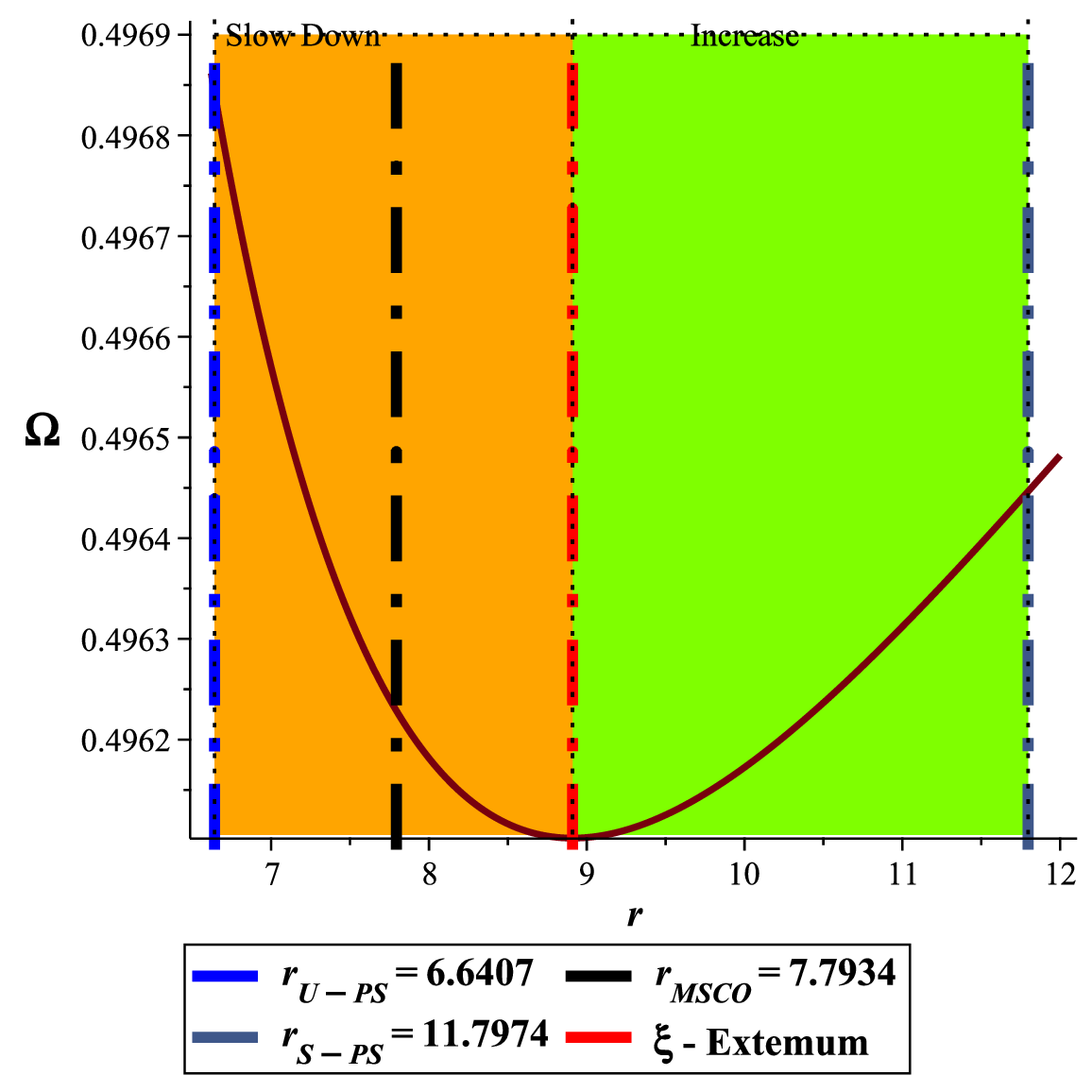}
 \label{6a}}
 \caption{\small{Angular velocity VS r with $l = 2, \beta = 1.5, c = 1, C_{1} = -1, C_{2} = 1, M = 0.8, q = 2, \alpha = 0.01, m = 0.5$ for  M-EGB-Massive BH }}
 \label{6}
\end{center}
\end{figure}
As seen in Fig. (\ref{6}), the answer is affirmative. This phenomenon is also observable in this model. Therefore, it can be concluded that in non-rotating models, the emergence of a stable photon sphere beyond the event horizon could be one of the conditions leading to the manifestation of this phenomenon.\\ However, a very important question that needs to be addressed is whether the existence of a stable photon sphere is the only necessary condition or if there are other conditions as well. Is this phenomenon truly absent in other black hole models and exclusive to these few specific models?
\\To address this question, let us take a closer look at the function $\xi$. As clearly seen in Eq. (\ref{(21)}), the angular velocity depends on the energy (E) and angular momentum (L). These quantities, in turn, depend on the real and positive nature of the function $\xi$. In other words, at (r) values where the function $\xi$ is negative, the energy and angular momentum become imaginary and lack physical interpretation. With some mathematical calculations, the relationship between $\xi$ and the metric coefficients can be easily derived. Thus, we have \cite{7}:
\begin{equation}\label{(22)}
\xi_{\pm}=-\Omega_{\pm}^{2} g_{\varphi \varphi}-g_{\mathit{tt}}.
\end{equation}
Given the metric form in Eq. (\ref{(3)}), the above relationship can be rewritten as follows:
\begin{equation}\label{(23)}
\Omega_{\pm}=\sqrt{\frac{f \! \left(r \right)-\xi_{\pm}}{r^{2}}}.
\end{equation}
A closer look at Figure (\ref{6}) shows one interesting aspect of this diagram: the determining role of the parameter $\xi$. As shown in the figure, even after passing the MSCO and entering the STCOs, the increase in velocity has not yet occurred. In this study, and according to Eq. (\ref{(23)}), it can be seen that, given the nature of f(r), and as long as $\xi$ is increasing relative to the radius, the velocity will generally decrease. The transition to an increasing velocity trend only occurs when $\xi$ starts to decrease, which we have indicated with a red dashed line in our graph.
\\In response to the question of why the Aschenbach phenomenon is not typically observable in most models, it should be noted that in gravitational models with usual black hole behavior, two fundamental points are evident.\\Firstly, when examining the $\xi$ function of these black holes, we find that this function always exhibits an increasing behavior in permissible regions, and if a region with decreasing behavior is found, it is in regions not permissible for the presence of TCOs, as shown in Fig. (\ref{7})\cite{12,13}.
Secondly, stable photon spheres in these models, although present, are located behind the event horizon. In other words, by applying the Weak Cosmic Censorship Conjecture (WCCC) and creating an event horizon, we effectively eliminate and ignore their effects. This means that most of these models, in their horizonless or naked singularity form, can clearly and easily demonstrate this phenomenon, as shown in Fig. (\ref{8}).
As can be seen in both parts of Fig. (\ref{8}), when the model is in the form of a naked singularity, both the S-PS (Stable Photon Sphere) appears, and in its vicinity, the parameter $\xi$ decreases, which implies an increase in $\Omega$.
\section{Conclusions}
In this paper, our aim was to study the Aschenbach effect in static or non-rotating structures. The significance of studying such a phenomenon can be examined from various perspectives. The Aschenbach effect can serve as additional evidence supporting the validity of general relativity through the study of black hole physics. In classical mechanics, the angular velocity of an object in a stable circular orbit decreases with increasing radius.  However, in general relativity, a rotating black hole drags spacetime around it due to its spin. This effect, known as frame-dragging, influences the motion of nearby objects, causing them to move in a more complex manner compared to classical mechanics, and the observed increase in angular velocity associated with the Aschenbach effect could also be evidence of this claim. 
Additionally, we know that any change in the velocity of particle motion implies a change in their energy. In the context of black hole structures, this change in velocity can lead to a chain of energy changes in a region around the black hole, resulting in observable visual evidence in the form of radiation. Specifically, the material in the disk emits X-rays at specific frequencies as it spirals inward due to the intense gravitational pull of the black hole. The increase in the angular velocity of the material at certain radii around the black hole can lead to distinct changes in quasi-periodic oscillations (QPO) frequencies.  Consequently, QPOs are one of the phenomena that the Aschenbach effect can influence.
Also, we know that when a small object spirals into a very massive black hole, it emits gravitational waves. These waves carry information about the dynamics of the system, including the behavior of matter near the black hole's event horizon. Another intriguing possibility for observing the Aschenbach effect could be through the detection of gravitational waves from extreme mass ratio inspirals (EMRIs) by the future LISA (Laser Interferometer Space Antenna) mission \cite{21}.\\
Of course, other evidence can also be cited. For instance, when high-resolution observations of rotational matter near the event horizon of supermassive black holes become available, this effect is potentially observable. Similar to those recently discovered at the Galactic center\cite{22,23}, or If the spin of a supermassive black hole is near-external (i.e. if the spin parameter of the black hole "$a> 0.9953$") the Aschenbach effect can be observed for co-rotating matter through the corresponding change of the radiation flux at different radial positions of the accretion disk or the flares \cite{18}.\\ 
However, all the aforementioned points apply to rotating black holes. Since this phenomenon appears to be specific to rotating black holes, its manifestation in non-rotating models could be highly challenging.\\ Initial investigation into the possibility of such a phenomenon in static black holes \cite{4} motivated us to, firstly, seek further evidence for the emergence of this effect by examining various models, and secondly, if black hole models exhibiting this property are found,  identifying the common mechanisms that lead to the Aschenbach effect in static black holes. Accordingly, we examined various black hole models with linear and nonlinear field structures (with examples included in this paper).
Among these models, we found that the M-EGB-Massive model could be a suitable candidate for demonstrating this phenomenon. We understood that the common feature of this model with two other studied models \cite{4,5} is the appearance of a stable photon sphere outside the event horizon. In other words, it seems that what causes acceleration and increased velocity for particles in non-rotating models is the emergence of a stable minimum in the studied spacetime.
Additionally, in comparing the models, we found that the  $\xi$ parameter also plays an interesting and significant role in this phenomenon. Specifically, as long as this function is increasing, the velocity decreases, and when it starts to decrease, this declining trend leads to an increase in velocity. This behavior pattern of the $\xi$ function, along with the fact that in black holes with usual behavior, only unstable photon spheres are displayed beyond the event horizon, prevents us from observing the Aschenbach phenomenon in these black hole models.
In fact, this effect also occurs in these models, but since we hide the stable photon sphere behind the event horizon by applying the Weak Cosmic Censorship Conjecture (WCCC) and creating an event horizon, the phenomenon is also hidden. In other words, in these models, when the model appears in the form of a naked singularity and the stable photon sphere becomes visible, this phenomenon can be clearly observed.\\
The fact that static black holes can also accelerate and alter the velocity of particles through the creation of a local minimum potential or a stable photon sphere is quite fascinating. This phenomenon suggests a uniform behavioral pattern for black holes. Additionally, as discussed in the first part of the conclusion regarding rotating black holes,  existing such a phenomenon in static black holes not only reinforces the theory of general relativity, but also paves the way for experimentally observation and  studying emissions from this kind of black hole  in the future.  
Finally, as a motivation for further research, it is worth noting that if a static black hole can mimic Aschenbach's behavior through a local minimum, 
then under certain conditions, such as the combination with known linear and nonlinear fields or gravitational corrections, a rotating black hole might also exhibit such a minimum beyond the event horizon. The important question will be what important implications could the emergence of this stable photon sphere have for the Aschenbach effect in rotating black holes?

\end{document}